# Observation of Interfacial Antiferromagnetic Coupling between Magnetic Topological Insulator and Antiferromagnetic Insulator

Fei Wang[1,2][*], Di Xiao[1][*], Wei Yuan[3,4][*], Jue Jiang[1], Yi-Fan Zhao[1], Ling Zhang[1], Yunyan Yao[3], Baojuan Dong[2], Wei Liu[2], Zhidong Zhang[2], Chaoxing Liu[1], Jing Shi[4], Wei Han[3,5], Moses H. W. Chan[1], Nitin Samarth[1], and Cui-Zu Chang[1]

[1]Department of Physics, Pennsylvania State University, University Park, Pennsylvania 16802, USA

[2]Shenyang National Laboratory for Materials Science, Institute of Metal Research, Chinese Academy of Sciences, Shenyang 110016, China

[3]International Center for Quantum Materials, School of Physics, Peking University, Beijing 100871, China

[4]Department of Physics, University of California, Riverside, California 92521, USA

[5]Collaborative Innovation Center of Quantum Matter, Beijing 100871, China

*These authors contributed equally to this work.

Corresponding authors: cxc955@psu.edu (C.Z.C.).

**Inducing magnetic orders in a topological insulator (TI) to break its time reversal symmetry has been predicted to reveal many exotic topological quantum phenomena. The manipulation of magnetic orders in a TI layer can play a key role in harnessing these quantum phenomena towards technological applications. Here we fabricated a thin magnetic TI film on an antiferromagnetic (AFM) insulator $Cr_2O_3$ layer**

**and found that the magnetic moments of the magnetic TI layer and the surface spins of the $Cr_2O_3$ layers favor interfacial AFM coupling. Field cooling studies show a crossover from negative to positive exchange bias clarifying the competition between the interfacial AFM coupling energy and the Zeeman energy in the AFM insulator layer. The interfacial exchange coupling also enhances the Curie temperature of the magnetic TI layer. The unique interfacial AFM alignment in magnetic TI on AFM insulator heterostructures opens a new route toward manipulating the interplay between topological states and magnetic orders in spin-engineered heterostructures, facilitating the exploration of proof-of-concept TI-based spintronic and electronic devices with multi-functionality and low power consumption.**

Topological insulator (TI), a material in which the interior is insulating but the electrons can travel without resistance along its surface/edge conducting channels, has radically changed the research landscape of condensed matter physics and material science in the past decade [1,2]. The nontrivial Dirac surface/edge states of a TI are induced by the strong spin-orbit coupling of the material and thus protected by time-reversal symmetry (TRS) [3-6]. Breaking the TRS of a TI with a magnetic perturbation can lead to a variety of exotic quantum phenomena such as the quantum anomalous Hall (QAH) effect [7-9], topological magnetoelectric effect [8,10], and image magnetic monopole [11]. The QAH effect has been experimentally demonstrated in magnetically doped TI thin films, specifically Cr- and/or V-doped $(Bi,Sb)_2Te_3$ [12,13]. To date, the critical temperature of the QAH state ($T_{QAH}$) which we define as the temperature below which the quantized Hall resistance is realized, in magnetically doped TI films is still ~1K [12-15]. A low $T_{QAH}$ impedes both the exploration of fundamental physics and meaningful technological applications based on this exotic phenomenon. A direct route to increase $T_{QAH}$ is to increase the magnetic doping level in the TI film to enhance the Curie temperature ($T_C$). However, this process invariably degrades the quality of TI films and can even make it become trivial [16]. An alternative approach to induce the ferromagnetic (FM) order in TI but keep its nontrivial property is called for.

FM order can also be introduced into a TI layer through proximity to an FM insulator layer. By not introducing magnetic ions into the TI, the sample quality, in particular, the carrier mobility is expected to be much higher [17]. Experimental efforts along this line have demonstrated proximity induced interfacial magnetization in TIs with a few FM insulators, including EuS [18,19], GdN [20], $BaFe_{12}O_{19}$ [21], $Cr_2Ge_2Te_6$ [22] and ferrimagnet yttrium/thulium iron garnet (YIG/TIG) [23-26]. Since the magnetic proximity effect is a short-range magnetic exchange interaction, an antiferromagnetic (AFM) insulator layer with uncompensated surface spins could play the same role as FM insulators [27]. AFM insulators have a number of advantages compared with FM insulators, such as their insensitivity to perturbing magnetic fields, the high THz operating frequencies, and the negligible stray fields. These are attractive properties for spintronic applications [28-30]. Since the Néel temperature ($T_N$) of AFM insulator is usually well above the room temperature, it may be possible to induce a high Curie temperature ($T_C$) FM order in a TI. Recently, a transport cum neutron scattering experiment has found the interfacial spin texture modulation and the $T_C$ enhancement in magnetically doped TI on AFM metal CrSb heterostructures [31]. In view of the metallic property of CrSb, it is not possible to single out the transport property of the TI layer from that of CrSb. Therefore, an insulating AFM substrate (i.e., AFM insulator) is a better candidate to induce magnetic orders in the TI layer. We know of no study to date on such an experimental system.

In this *Letter*, we grew AFM insulator $Cr_2O_3$ layers with different thicknesses on heat-treated sapphire (0001) substrate to be followed with 4 quintuple layers (QL) magnetic TI $Sb_{1.8}Cr_{0.2}Te_3$ layer to form $Sb_{1.8}Cr_{0.2}Te_3/Cr_2O_3$ heterostructures. We demonstrated that when the thickness of the $Cr_2O_3$ layer (*m*) is ≤ 3-unit cell (UC, 1UC ~1.36 nm), the anomalous Hall (AH) hysteresis loops of the magnetic TI layer, measured by systematically changing the magnitude of the magnetic field ($\mu_0 H_{CF}$) employed for field cooling, show a crossover from negative to positive shifts (i.e., exchange bias) along the magnetic field axis. This observation indicates an interfacial AFM coupling between the magnetic moments of the magnetic TI layers and the surface spins of the AFM $Cr_2O_3$ layer. The crossover of negative to positive exchange bias disappears for *m* ≥ 4UC because the $T_N$ of the thicker $Cr_2O_3$ layer is higher than the $T_C$ of the magnetic TI layer and this makes the surface spins of the AFM layer aligned randomly and smears out the appearance of the exchange bias. Upon further increase of the thickness of the $Cr_2O_3$ layer, the $T_C$ of $Sb_{1.8}Cr_{0.2}Te_3$ layers is progressively enhanced from ~39K without the $Cr_2O_3$ layer (see Supporting Materials) to ~50K for $Cr_2O_3$ layer thicker than 14UC. The $T_C$ enhancement also confirms the existence of the interfacial exchange coupling between magnetic TI and the $Cr_2O_3$ layers.

Bulk $Cr_2O_3$ is a well-known AFM insulator with a $T_N$ of 307K [32], whose linear magnetoelectric property has been used in voltage-controlled spintronic devices [33-35]. The spins along the (0001) direction in the intralayer are FM aligned, while the spins of the adjacent layers are AFM coupled (**Fig. 1a**). It is known that the $T_C$ ($T_N$) of the FM (AFM) films due to the finite size effect is usually lower than its bulk value [36,37]. Therefore, the $T_N$ of AFM $Cr_2O_3$ films can be controlled by varying $m$. The $Cr_2O_3$ layers with different $m$ (from 1 to 35 UC) were deposited at 500°C by pulsed laser deposition (PLD) on heat-treated sapphire (0001) substrates [38,39]. The growth process was monitored by *in-situ* reflection high-energy electron diffraction (RHEED). The sharp and streaky "1×1" patterns indicate highly-ordered $Cr_2O_3$ films with atomically flat surfaces (**Fig. S1**). The high quality of the $Cr_2O_3$ films is also confirmed by atomic force microscopy and high-resolution X-ray diffraction (HR-XRD) measurements (**Figs. S2 and S3**).

The growth of the 4 QL $Sb_{1.8}Cr_{0.2}Te_3$ films on AFM $Cr_2O_3$ layers was carried out in a molecular beam epitaxy (MBE) chamber with a base pressure of $2\times10^{-10}$ mbar. During the growth of the magnetic TI film, the $Cr_2O_3$/sapphire substrate was maintained at ~240°C. This low MBE growth temperature inhibits the diffusion of the Cr atoms into the Cr-doped $Sb_2Te_3$ layer. The high quality of the 4QL $Sb_{1.8}Cr_{0.2}Te_3$ is confirmed by sharp "1×1" RHEED patterns, the smooth atomic force microscopy image that shows a roughness of ~0.9 nm over 5μm × 5μm area and the sharp (00$n$) peaks in the HR-XRD spectroscopy (**Figs. S4 to S6**). To avoid possible contamination, a 10 nm thick Te layer was deposited at room temperature on the magnetic TI film prior to the removal of the heterostructure samples from the MBE chamber for *ex-situ* measurements.

**Figure 1b** shows the cross-sectional scanning transmission electron microscopy (STEM) image of the 10nm Te capped 4QL $Sb_{1.8}Cr_{0.2}Te_3$/35UC $Cr_2O_3$ heterostructure grown on a sapphire substrate and the corresponding energy dispersive spectroscopy (EDS) mappings of Al, Cr, Sb, and Te. Al and Cr are found residing at the sapphire substrate and Cr$_2$O$_3$ layers, respectively, while Sb and Te are found in the Cr-doped TI and Te capping layers. The absence of the Cr signal in Cr-doped Sb$_2$Te$_3$ is due to the low concentration of Cr in the TI layer. The trace Sb signal showing in the Te capping layer is a result of their neighboring peak positions in the X-ray spectrum. More TEM and EDS results on other two heterostructures (7 QL Sb$_{1.8}$Cr$_{0.2}$Te$_3$/3UC Cr$_2$O$_3$ and 4QL Sb$_{1.8}$Cr$_{0.2}$Te$_3$/14UC Cr$_2$O$_3$) are shown in Supporting Materials (**Fig. S7**).

The 10nm Te capped 4QL Sb$_{1.8}$Cr$_{0.2}$Te$_3$/$m$ UC Cr$_2$O$_3$ heterostructure samples were scratched into a Hall bar geometry using a computer-controlled probe station [10]. Transport studies were carried out in a Physical Property Measurement System (Quantum Design, 2 K, 9 T) with an external magnetic field perpendicular to the film. The excitation current is 1 μA. All samples were field cooled from 320K, which is above $T_N$ of bulk Cr$_2$O$_3$ (~307K) [32], at different specific external magnetic fields μ$_0H_{CF}$ to the target measurement temperatures. The field cooling process eliminates any possible spontaneous and random alignments of the AFM order in the Cr$_2$O$_3$ layer. **Figure 2a** shows the magnetic field (μ$_0H$) dependence of the Hall resistance ($\rho_{yx}$) at 2K of the 4QL Sb$_{1.8}$Cr$_{0.2}$Te$_3$/3UC Cr$_2$O$_3$ heterostructure field cooled at μ$_0H_{CF}$ = 0.05, 0.3, 0.8, 2 and 7T. The blue (red) curves correspond to $\rho_{yx}$ measured while sweeping μ$_0H$ downward from 0.2T to -0.2T (upward from -0.2T to 0.2T). The small μ$_0H$ sweep range is chosen to make sure the spin orders of the AFM layers induced by the field cooling procedure are preserved. The nearly square $\rho_{yx}$ hysteresis loops confirm a well-defined FM order with perpendicular magnetic anisotropy in these films [12,40]. The $\rho_{yx}$ loop for the sample prepared at μ$_0H_{CF}$ = 0.05T shows a negative exchange bias, i.e., the left coercive field |μ$_0H_{cL}$| is larger than the right coercive field μ$_0H_{cR}$ thus indicating the presence

of a unidirectional magnetic exchange anisotropy across the $Sb_{1.8}Cr_{0.2}Te_3/Cr_2O_3$ interface. The $\mu_0H_{CF} = 0.3T$ loop still shows the negative exchange bias, but the magnitude of the shift is reduced. The $\mu_0H_{CF} = 0.8T$ loop is nearly symmetric, aka, $|\mu_0H_{cL}|=\mu_0H_{cR}$. Upon further increase of $\mu_0H_{CF}$ to 2T, the negative exchange bias (i.e., $|\mu_0H_{cL}|>\mu_0H_{cR}$) is replaced by positive exchange bias (i.e., $|\mu_0H_{cL}|<\mu_0H_{cR}$). The positive exchange bias becomes more pronounced if the sample is cooled under $\mu_0H_{CF}=7T$. The observation of a crossover from negative to positive exchange bias indicates that opposite AFM domain states in the $Cr_2O_3$ layer were established under different $\mu_0H_{CF}$.

The Hall traces of the 4QL $Sb_{1.8}Cr_{0.2}Te_3$/2UC $Cr_2O_3$ and 4QL $Sb_{1.8}Cr_{0.2}Te_3$/1UC $Cr_2O_3$ are shown respectively in **Figs. 2b** and **2c**. These two samples also display negative exchange bias under low $\mu_0H_{CF}$ and positive exchange bias under high $\mu_0H_{CF}$. **Figures 2a** to **2c** show that the $\mu_0H_{CF}$ at which a symmetric $\rho_{yx}$ loop is found monotonically decreases with decreasing thickness of the $Cr_2O_3$ layers. The magnitude of the exchange bias field ($\mu_0H_E$) is defined as $(\mu_0H_{cR}-|\mu_0H_{cL}|)/2$. The $\mu_0H_E$s as a function of $\mu_0H_{CF}$ at $T=2K$ for the 4 QL $Sb_{1.8}Cr_{0.2}Te_3/m$ UC $Cr_2O_3$ heterostructures ($m =1\sim4$) are summarized in **Fig. 2d**. We note that for $m\leqslant 3$ UC, the $\mu_0H_E$ changes from a negative value through zero to a positive value when the $\mu_0H_{CF}$ is discretely increased from 0.05T to 7T. The critical $\mu_0H_{CF}$ with $\mu_0H_E =0$ is labeled as $\mu_0H_{CF}^0$. The $\mu_0H_{CF}^0$ are 0.8T, 0.7T and 0.3T for 3UC, 2UC, and 1UC $Cr_2O_3$ layers, respectively (Inset of **Fig. 2d**). However, for $m = 4$ UC, the $\rho_{yx}$ loop is always symmetric, i.e., showing a negligible shift under all $\mu_0H_{CF}$, indicating the suppression of an exchange bias effect (**Fig. 2d** and **Fig. S9**).

Magnetic hysteresis loop shifts as a result of the interfacial exchange interaction have been observed in many systems with FM-AFM interfaces [41]. The AFM layer usually plays the role of a pinning layer, whose alignment direction determines the shift direction of the hysteresis loop. In most systems, the shift of the hysteresis loop is along the direction opposite to $\mu_0H_{CF}$, i.e., showing negative exchange bias, which is an evidence for the interfacial FM coupling [41]. In our Cr-doped $Sb_2Te_3$ on $Cr_2O_3$ heterostructures, we observed the crossover of the negative to positive exchange bias by systematically increasing $\mu_0H_{CF}$. The observation of the positive exchange bias under high $\mu_0H_{CF}$ is a necessary condition for the interfacial AFM coupling between the magnetic moments of the FM layer and the surface spins of the AFM layer [41-44]. More details will be shown below.

Next, we focus on the study of the evolution of positive $\mu_0H_E$ of 4QL $Sb_{1.8}Cr_{0.2}Te_3$ on $Cr_2O_3$ of 3, 2 and 1UC field-cooled with $\mu_0H_{CF}$ =7T. The $\rho_{yx}$ hysteresis loops of these heterostructures at temperatures from 2 to 40K are shown respectively in **Figs. 3a** to **3c**. These loops show that $\mu_0H_{cR}$, $|\mu_0H_{cL}|$, and the positive $\mu_0H_E$, all monotonically decrease with increasing temperature. **Figure 3d** shows $\mu_0H_E$ as a function of temperature for these heterostructures. The blocking temperature $T_B$, i.e., the temperature above which $\mu_0H_E$ vanishes, increases with increasing $m$ of the $Cr_2O_3$ layers. The $T_B$s for the heterostructures with 1UC, 2UC, and 3UC $Cr_2O_3$ layers are 15K, 25K, and 35K, respectively. The higher $T_B$ in the heterostructures with thicker $Cr_2O_3$ layers suggests the $T_N$ is higher in thicker $Cr_2O_3$ layers.

To achieve the exchange bias effect in the systems with the FM-AFM interface, the $T_N$ of the AFM layer must be lower than the $T_C$ of the FM layer. Therefore, for $m \leqslant$ 3UC, the $T_N$ of $Cr_2O_3$ layer must be lower than the $T_C \sim$ 38K of 4QL $Sb_{1.8}Cr_{0.2}Te_3$ (**Fig. S13**). We now explain why the observation of the crossover from the negative to positive exchange bias demonstrates the magnetic moments of the Cr-doped $Sb_2Te_3$ and $Cr_2O_3$ layers are AFM coupled. Since $T_C \geq T_N$, during the cooling of the samples under a positive $\mu_0H_{CF}$, the temperature first arrives at $T=T_C$, the magnetic moments of the 4QL $Sb_{1.8}Cr_{0.2}Te_3$ layer are consequently aligned upwards as shown in **Figs. 4a** and **4b**. With further cooling to $T = T_N$, the AFM order is formed in $Cr_2O_3$ layers. If the AFM and FM layers favor the interfacial AFM coupling, the spins of the aligned $Sb_{1.8}Cr_{0.2}Te_3$ layer exert an effective field to force the surface spins of the $Cr_2O_3$ layer to be oppositely aligned, i.e., pointing downwards. The AFM coupling energy between the AFM and FM layers is $J_E S_{AFM} S_{FM}$, where $J_E$ is the exchange coupling strength between the surface spins of $Cr_2O_3$ ($S_{AFM}$) and $Sb_{1.8}Cr_{0.2}Te_3$ ($S_{MTI}$) layers. In addition, the Zeeman energy in the AFM layer under the external magnetic field $\mu_0H_{CF}$ is $\mu_0H_{CF}M_{AFM}$, where $M_{AFM}$ is the surface magnetization of the AFM $Cr_2O_3$ layer. This Zeeman energy forces the surface spins of the $Cr_2O_3$ layer to be aligned along the direction of the external magnetic field [42,44]. The alignment between AFM and FM layers is determined by the competition between the AFM coupling energy $J_E S_{AFM} S_{FM}$ and the Zeeman energy in the AFM layer $\mu_0H_{CF}M_{AFM}$.

When the sample is field-cooled under a low $\mu_0H_{CF}$, $\mu_0H_{CF}M_{AFM} < J_E S_{AFM} S_{FM}$ (**Fig. 4a**), the AFM coupling energy $J_E S_{AFM} S_{FM}$ is then dominant and, as a consequence, the positive magnetization of the Cr-doped $Sb_2Te_3$ will force the top surface spins of the $Cr_2O_3$ layer to point downwards. With a high $\mu_0H_{CF}$, $\mu_0H_{CF}M_{AFM} > J_E S_{AFM} S_{FM}$ (**Fig. 4b**), the Zeeman energy in $Cr_2O_3$ layer $\mu_0H_{CF}M_{AFM}$ is dominant and makes the top surface spins of the $Cr_2O_3$ layer to point upwards. To compensate for the AFM coupling, the orientation of the surface spins in the $Cr_2O_3$ layer will be slightly tilted. Since the magnetic field sweep range ±0.2T is much lower than the magnetic field at which the spin-flop transition, reported to occur at a few Teslas [39,45], the adjacent magnetic layers in the $Cr_2O_3$ layer are still AFM ordered but are slightly tilted (**Fig. 4b**). With a critical value of $\mu_0H_{CF}$ (i.e., $\mu_0H_{CF}^0$), $\mu_0H_{CF}^0 M_{AFM}=J_ES_{AFM}S_{FM}$, the $Cr_2O_3$ layer is in a randomly distributed multi-domain state, and thus $\mu_0H_E=0$, and gives rise to the symmetric hysteresis loop. $\mu_0H_{CF}^0$ increases with $m$ as a result of the $J_E$ enhancement, which is induced by the higher $T_N$ in thicker $Cr_2O_3$ layers [44].

The surface spin configuration of the $Cr_2O_3$ layer at low temperature is locked by its AFM magnetic structure during the field cooling process. When the external $\mu_0H$ is swept in the ±0.2T range the FM spins of the 4QL $Sb_{1.8}Cr_{0.2}Te_3$ layer with a $\mu_0H_c$ of ~0.1T can be switched by the external magnetic field while the AFM spins of the $Cr_2O_3$ layer are unchanged [39,45]. The existence of the AFM coupling energy $J_ES_{AFM}S_{FM}$ across Cr-doped $Sb_2Te_3$ and $Cr_2O_3$ interface will cause the shift of the FM hysteresis loop of the 4QL $Sb_{1.8}Cr_{0.2}Te_3$ layer. For samples field cooled with a low $\mu_0H_{CF}$, the top surface spins of the $Cr_2O_3$ layer are locked downwards when $\mu_0H$ is swept at $T$=2K and the magnetic moments of the 4QL $Sb_{1.8}Cr_{0.2}Te_3$ layer are pointing upwards for $\mu_0H$= 0.2T (Step I in **Fig. 4c**). When $\mu_0H$ is swept downward, near the left coercive field $\mu_0H_{cL}$, the upward magnetic domains of the 4QL $Sb_{1.8}Cr_{0.2}Te_3$ layer is reversed. Due to the interfacial AFM coupling, the spins of $Sb_{1.8}Cr_{0.2}Te_3$ are energetically more favorable to point upwards to retain the antiparallel alignment. Therefore, to overcome the interfacial AFM coupling energy, a magnetic field $\mu_0H$ larger than $\mu_0H_c$ is needed to reverse the magnetic domains from upwards to downward, i.e., $|\mu_0H_{cL}|> \mu_0H_c$ (Step II in **Fig. 4c**). When $\mu_0H$ is swept back from -0.2T, the presence of the interfacial AFM coupling helps the magnetization reversal of 4QL Sb$_{1.8}$Cr$_{0.2}$Te$_3$ layer from parallel to antiparallel alignments, so the right coercive field $\mu_0 H_{cR} < \mu_0 H_c$ (Step IV in **Fig. 4c**). This explains why negative exchange bias, i.e., $|\mu_0 H_{cL}| > \mu_0 H_{cR}$, is observed under a low $\mu_0 H_{CF}$.

When the sample is field-cooled under a high $\mu_0 H_{CF}$, the magnetic moments at $\mu_0 H = 0.2$T of the Cr-doped Sb$_2$Te$_3$ are still pointing upwards, but the top surface spins of the Cr$_2$O$_3$ layer are also pointing upwards but slightly tilted (Step I in **Fig. 4d**). When $\mu_0 H$ is swept downward, the existence of the AFM coupling will favor the magnetization reversal of 4QL Sb$_{1.8}$Cr$_{0.2}$Te$_3$ layer from upwards to downwards, so $|\mu_0 H_{cL}| < \mu_0 H_c$ (Step II in **Fig. 4d**). When $\mu_0 H$ is swept back from -0.2T, the interfacial AFM coupling will impede the magnetization reversal of 4QL Sb$_{1.8}$Cr$_{0.2}$Te$_3$ layer from downwards to upwards, so $\mu_0 H_{cR} > \mu_0 H_c$. This corresponds to the positive exchange bias, i.e., $|\mu_0 H_{cL}| < \mu_0 H_{cR}$, observed for the samples field cooled with a high $\mu_0 H_{CF}$.

We have also systematically studied the $T_C$ of the 4QL Sb$_{1.8}$Cr$_{0.2}$Te$_3$ layer for $m \geq 4$. The $T_C$ of 4QL Sb$_{1.8}$Cr$_{0.2}$Te$_3$ layer in all heterostructures was determined by the Arrott-plots (**Fig. S11**) [46]. The $m$ dependence of $T_C$ is summarized in **Fig. S12**. For $m \leq 4$UC, $T_C$ is ~ 39K, consistent with the $T_C$ of 4QL Sb$_{1.8}$Cr$_{0.2}$Te$_3$ film grown on the nonmagnetic SrTiO$_3$(111) substrate (**Fig. S13**). For $m \geq 5$UC, the $T_C$ starts to increase and saturates near 50K for $m \geq 14$UC. The $T_C$ enhancement of 4QL Sb$_{1.8}$Cr$_{0.2}$Te$_3$ grown on the thicker Cr$_2$O$_3$ films further demonstrate the existence of the interfacial exchange coupling.

To summarize, we demonstrated the tuning of the exchange bias in a given magnetic TI on AFM insulator heterostructure from negative to positive values by field cooling with different $\mu_0 H_{CF}$. This is made possible by the interfacial AFM coupling between the FM aligned spins of the magnetic TI layer and the surface spins of the AFM insulator. This efficient tuning of the exchange bias provides a new route to effectively manipulate the magnetic spins of the TI layer. Our findings, when combined with the linear magnetoelectric effect of $Cr_2O_3$ [33-35], could facilitate the development of proof-of-concept electric field-controlled TI-based energy-efficient spintronic devices.

## Acknowledgments

The authors would like to thank B. H. Yan and X. D. Xu for the helpful discussions. F. W., Y. Z. and C. Z. C. acknowledge the support from ARO Young Investigator Program Award (W911NF1810198) and the Alfred P. Sloan Research Fellowship. D. X. and N. S. acknowledge the support from the Penn State Two-Dimensional Crystal Consortium-Materials Innovation Platform (2DCC-MIP) under NSF grant DMR-1539916 and the Office of Naval Research (N00014-15-1-2370). C. X. L. acknowledges the support from the Office of Naval Research (N00014-15-1-2675). F. W., W. L., and Z. D. Z. acknowledge the support from the State Key Program of Research and Development of China (2017YFA0206302), the National Natural Science Foundation of China (NSFC) (51590883, 51331006 and 51771198) and the Key Research Program of Chinese Academy of Sciences (KJZD-EW-M05-3). J. S. acknowledges the support from DOE grant (DE-FG02-07ER46351). W. H. acknowledges the support from the National Basic Research Programs of China (2015CB921104) and the National Natural Science Foundation of China (NSFC) (11574006). Support for the electrical transport measurements and data analysis is provided by the DOE grant (DE-SC0019064).

**Figures and figure captions:**

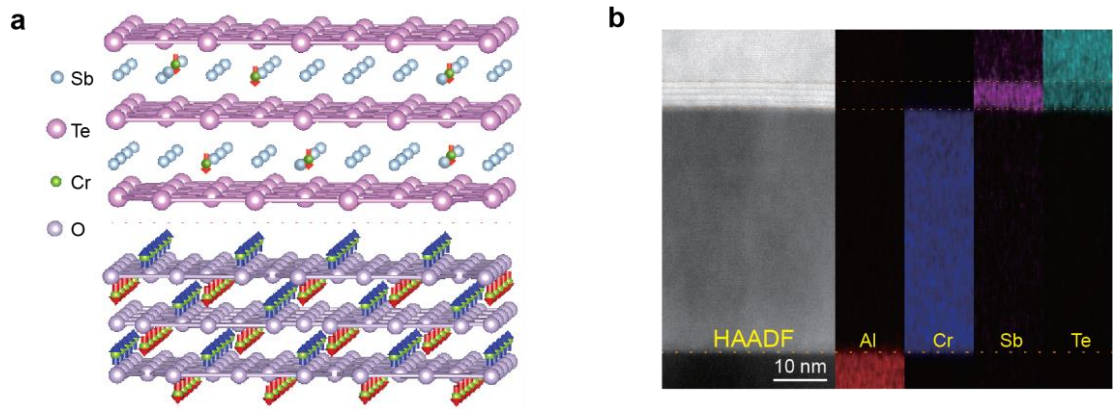

**Figure 1| The magnetic TI Cr-doped Sb$_2$Te$_3$/AFM Cr$_2$O$_3$ heterostructure.** (a) Schematic atomic structure of the Cr-doped Sb$_2$Te$_3$/Cr$_2$O$_3$ heterostructure. The magnetic moments of Cr-doped Sb$_2$Te$_3$ and the surface spins of the Cr$_2$O$_3$ layer are AFM aligned. (b) STEM image of the Te layer-capped 4 QL Sb$_{1.8}$Cr$_{0.2}$Te$_3$/35UC Cr$_2$O$_3$ heterostructure grown on a sapphire substrate, accompanied by an EDS map of Al, Cr, Sb, and Te of the sample.

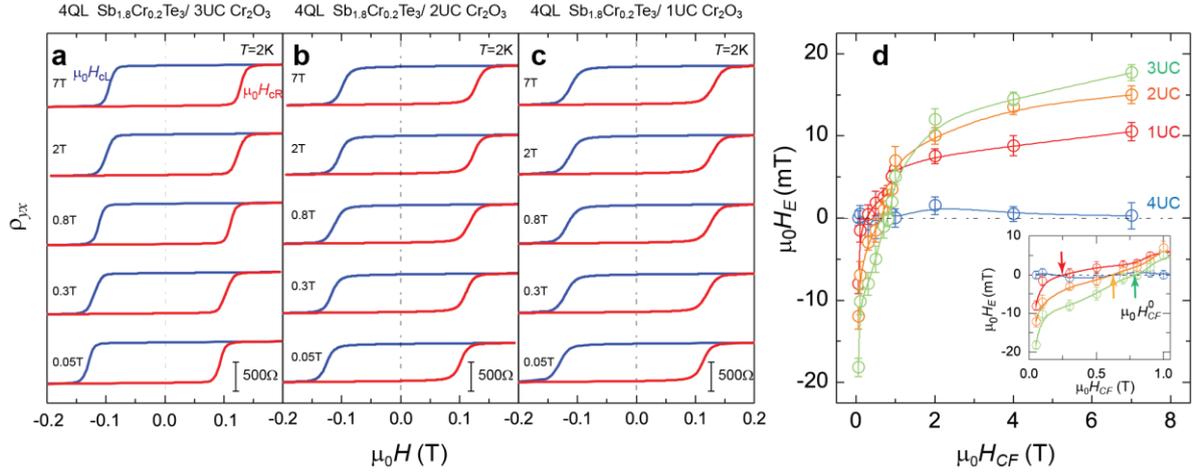

**Figure 2| AFM interfacial coupling between Cr-doped $Sb_2Te_3$ and $Cr_2O_3$ layers as revealed by the observation of crossover from negative to positive exchange bias.** (a, b, c) The Hall resistance $\rho_{yx}$ at $T$ = 2K of 4 QL $Sb_{1.8}Cr_{0.2}Te_3$ layer grown on 3 UC (a), 2 UC (b), 1 UC (c) $Cr_2O_3$ layers field cooled with $\mu_0 H_{CF}$=0.05, 0.3, 0.8, 2, and 7T. (d) The exchange bias $\mu_0 H_E$ as a function of $\mu_0 H_{CF}$ at $T$ =2K for 4 QL $Sb_{1.8}Cr_{0.2}Te_3$ grown on 1~4 UC $Cr_2O_3$ layers. Note that $\mu_0 H_E$ is negligible in the 4 QL $Sb_{1.8}Cr_{0.2}Te_3$/4 UC $Cr_2O_3$ heterostructure since the $T_N$ of 4 UC $Cr_2O_3$ layer is higher than $T_C$ of 4 QL $Sb_{1.8}Cr_{0.2}Te_3$ layer. The error bars reflect the standard deviations of the left and right coercive fields $\mu_0 H_{cL}$ and $\mu_0 H_{cR}$ used in the $\mu_0 H_E$ calculations. The inset shows the zoomed-in low $\mu_0 H_{CF}$ region. The horizontal intercepts denoted by the arrows become larger in heterostructures with the thicker $Cr_2O_3$ layer.

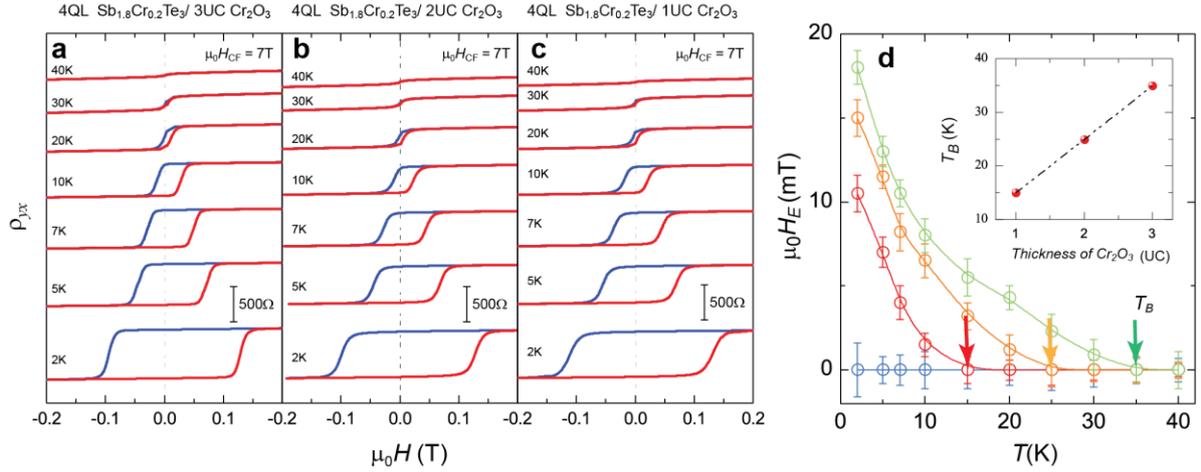

**Figure 3| Temperature dependence of the positive exchange bias in Cr-doped Sb$_2$Te$_3$/Cr$_2$O$_3$ heterostructures field cooled at μ$_0$H$_{CF}$=7T.** (a, b, c) The Hall resistance $\rho_{yx}$ at varying temperatures of 4 QL Sb$_{1.8}$Cr$_{0.2}$Te$_3$ layer on 3 UC (a), 2 UC (b), 1 UC (c) Cr$_2$O$_3$ layers. (d) Temperature dependence of the exchange bias μ$_0$H$_E$ in 4 QL Sb$_{1.8}$Cr$_{0.2}$Te$_3$ grown on 1~4 UC Cr$_2$O$_3$ heterostructures. The blocking temperature $T_B$, at which μ$_0$H$_E$=0, increases with increasing thickness of the Cr$_2$O$_3$ layer. The magnitude of the error bars reflects the standard deviations of the μ$_0$H$_{cL}$ and μ$_0$H$_{cR}$ used in calculating μ$_0$H$_E$. Inset shows $T_B$ as a function of the thickness of the Cr$_2$O$_3$ layer in the heterostructures.

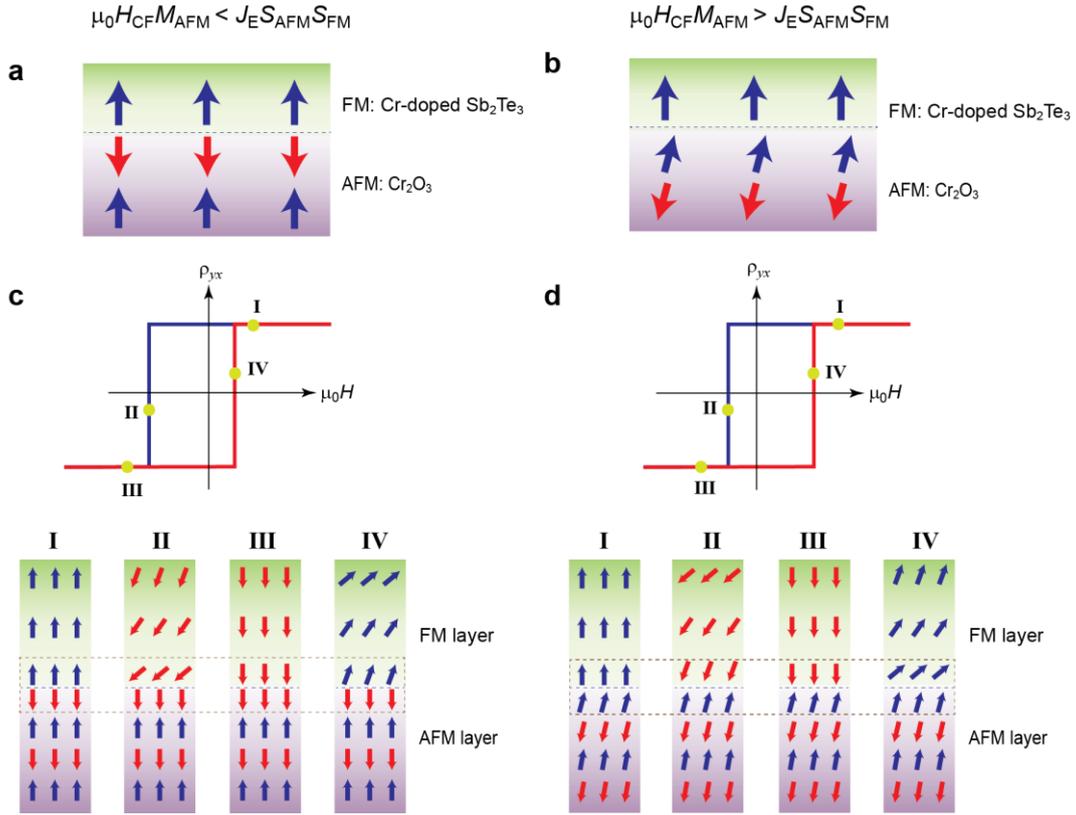

**Figure 4| A schematic of the model of interfacial AFM coupling between Cr-doped Sb$_2$Te$_3$ and Cr$_2$O$_3$ layers.** (a, b) The orientations of the magnetic moments in the Cr-doped Sb$_2$Te$_3$ and Cr$_2$O$_3$ layers field cooled under low (a) and high (b) $\mu_0 H_{CF}$. When the heterostructure is field cooled under a low $\mu_0 H_{CF}$ ($\mu_0 H_{CF} M_{AFM} < J_E S_{AFM} S_{FM}$), the spins of the Cr-doped TI and Cr$_2$O$_3$ layers are AFM aligned. For a high $\mu_0 H_{CF}$ ($\mu_0 H_{CF} M_{AFM} > J_E S_{AFM} S_{FM}$), the spins of the Cr-doped TI and Cr$_2$O$_3$ layers are forced to be FM aligned. (c) Negative exchange bias and the spin switching process of the Cr-doped TI layer prepared with a low $\mu_0 H_{CF}$. (d) Positive exchange bias and the spin switching process of the Cr-doped TI layer prepared with a high $\mu_0 H_{CF}$.